\documentclass{aa}
\usepackage{graphicx}
\usepackage{txfonts}

\newcommand{\Msun}{\ensuremath{\,{\rm M}_\odot}}                  
\newcommand{\Rsun}{\ensuremath{\,{\rm R}_\odot}}                  
\newcommand{\Lsun}{\ensuremath{\,{\rm L}_\odot}}                  
\newcommand{\lsun}{\ensuremath{L_\odot}}                          
\newcommand{\Teff}{\ensuremath{T_{\rm eff}}}                      
\newcommand{\Vsync}{\ensuremath{V_{\rm synch}}}                   
\newcommand{\EBV}{\ensuremath{E_{B-V}}}                           
\newcommand{\Eby}{\ensuremath{E_{b-y}}}                           
\newcommand{\kms}{\,km\,s$^{-1}$}                                 
\newcommand{\mc}[1]{\multicolumn{2}{c}{#1}}                       
\newcommand{\Mbolsun}{\ensuremath{M_{\rm bol \odot}}}             
\newcommand{\Mbol}{\ensuremath{M_{\rm bol}}}                      
\newcommand{\sci}[2]{{#1}\!\times10^{#2}}                         

\begin{document}
\title{Eclipsing binaries as standard candles:}
\subtitle{HD\,23642 and the distance to the Pleiades}
\titlerunning{HD\,23642 and the distance to the Pleiades}
\author{J.\ Southworth \and P.\ F.\ L.\ Maxted \and B.\ Smalley}
\offprints{J.\ Southworth}
\institute{Department of Physics and Chemistry, Keele University, Staffordshire, ST5 5BG, UK\\
           \email{jkt@astro.keele.ac.uk (JS), pflm@astro.keele.ac.uk (PFLM), bs@astro.keele.ac.uk (BS)}}
\date{Received ???? / Accepted ????}

\abstract{We present a reanalysis of the light curves of HD\,23642, a detached eclipsing binary star in the Pleiades open cluster, with emphasis on a detailed error analysis. We compare the masses and radii of the two stars to predictions of stellar evolutionary models and find that the metal and helium abundances of the Pleiades are approximately solar. We present a new method for finding distances to eclipsing binaries, of spectral types A to M, using the empirical calibrations of effective temperature versus surface brightness given by Kervella et al. (\cite{KTDS}). We use the calibration for $K$-filter surface brightness to determine a distance of $139.1 \pm 3.5$\,pc to HD\,23642 and the Pleiades. This distance is in excellent agreement with distances found from the use of theoretical and empirical bolometric corrections. We show that the determination of distance, both from the use of surface brightness relations and from the use of bolometric corrections, is more accurate and precise at infrared wavelengths than at optical wavelengths. The distance to HD\,23642 is consistent with that derived from photometric methods and Hubble Space Telecscope parallaxes, but is inconsistent with the distance measured using Hipparcos parallaxes of HD\,23642 and of other Pleiades stars.
}

\maketitle

\section{Introduction}                                                                                      \label{intro}

The \object{Pleiades} is a nearby, young open star cluster which is of fundamental importance to our understanding of stellar evolution and the cosmic distance scale. It has been exhaustively studied by many researchers and its distance and chemical composition were, until recent observations, considered to be well established. The distance derived from data obtained by the Hipparcos satellite, however, is in disagreement with traditional values, leading to claims that stellar evolutionary theory is much less reliable than previously thought.

The `long' distance scale of $132 \pm 3$\,pc was established by main sequence fitting analyses (e.g., Johnson \cite{johnson}; Meynet, Mermilliod \& Maeder \cite{MMM}). Recent parallax observations from terrestrial telescopes (Gatewood, de Jonge \& Han \cite{GDH}), and from the Hubble Space Telescope (Benedict et al. \cite{benedict}) are in good agreement with this distance. 

The astrometric binary \object{HD\,23850} (Atlas) was recently studied by Pan, Shao \& Kulkarni (\cite{PSK}) using the Palomar Testbed Interferometer (Colavita et al.\ \cite{CWH}). These authors did not have a spectroscopic orbit for HD\,23850 available to them, but were able to show that the distance to Atlas was greater than 127\,pc, and probably between 133 and 137\,pc. Zwahlen et al.\ (\cite{zwahlen}) have subsequently published a spectroscopic orbit and new interferometric measurements which, combined with the observations of Pan et al.\ (\cite{PSK}), give an entirely geometrical distance of $132 \pm 4$\,pc to HD\,23850.

A `short' distance scale of $120 \pm 3$\,pc (van Leeuwen \cite{leeuwen04}) has been found using trigonometrical parallaxes observed by the Hipparcos space satellite (Perryman et al. \cite{perryman}). This is in conflict with the traditional `long' distance scale for the Pleiades. In an attempt to explain this, van\,Leeuwen (1999) placed the main sequences of other nearby open clusters in the HR diagram using Hipparcos parallaxes, and found that five of the eight clusters have main sequences as faint as the Pleiades.

Castellani et al.\ (\cite{CDPT}) have shown that current theoretical stellar evolutionary models can fit the Pleiades main sequence if a low metal abundance of $Z = 0.012$ is adopted. However, Stello \& Nissen (\cite{SN}) used a metallicity-insensitive photometric technique to demonstrate that, if the Hipparcos parallaxes were correct, the main sequence Pleiades stars were implausibly fainter than their counterparts in the field. Also, Boesgaard \& Friel (\cite{BF}) have measured the iron abundance of the Pleiades to be approximately solar (${\rm [Fe/H]} = -0.034 \pm 0.024$) from high-resolution spectra of twelve F dwarfs in the cluster (further references can be found in Stauffer et al.\ \cite{stauffer}).

Narayanan \& Gould (\cite{NG}) have presented evidence that the Hipparcos parallaxes are correlated on angular scales of two to three degrees. They used a variant of the moving cluster method, based on radial velocities, to find a distance of $130 \pm 11$\,pc, in agreement with both the 'long' distance scale and the `short' Hipparcos distance (van Leeuwen \cite{leeuwen04}). Makarov (\cite{makarov}) has reanalysed the Hipparcos data, allowing for this suggested correlation, and found the Pleiades distance to be $129 \pm 3$\,pc. Until this result is confirmed, however, the `long' and `short' distance scales cannot yet be considered to be reconciled.

Munari et al. (\cite{M04} hereafter M04) studied the detached eclipsing binary \object{HD\,23642} and found a distance of $132 \pm 2$\,pc, in good agreement with the `long' distance scale. The method used by M04 is commonly used to find the distances to eclipsing binaries but depends on theoretical calculations to provide bolometric corrections. We have reanalysed the data of M04 (which these authors have graciously provided on the internet) to investigate alternative, empirical, methods of finding the distance to HD\,23642 and similar eclipsing binaries by the use of surface brightness relations.

\subsection{The eclipsing binary HD\,23642 in the Pleiades}                                                 \label{sec:HD23642}

Detached eclipsing binaries (dEBs) with double-lined spectra are one of the best sources of fundamental astrophysical data (Andersen \cite{andersen}) because their absolute masses and radii can be measured to accuracies better than 1\%. Such data can be used to provide a strict test of different stellar evolutionary models, and the distances to dEBs can be determined empirically to an accuracy of about 5\%. dEBs in open clusters are particularly useful because the age and chemical composition of the cluster can be combined with accurate values of the masses and radii of the dEB to provide an even more exacting test of theoretical models (Southworth, Maxted \& Smalley \cite{SMSa}, \cite{SMSb}).

HD\,23642 (Table~\ref{table:photdata}) was discovered to be a double-lined spectroscopic binary by Pearce (\cite{pearce}) and Abt (\cite{abt}), and both components have been found to display slight spectral peculiarities (Abt \& Levato \cite{abtlevato}). Torres (\cite{torres}) discovered shallow secondary eclipses in the Hipparcos photometric data of HD\,23642 and also presented an accurate spectroscopic orbit. M04 derived precise absolute masses and radii of both components from high-resolution spectra and complete $BV$ light curves. M04 found a distance of $131.9 \pm 2.1$\,pc, in disagreement with the Hipparcos parallax distance of $111 \pm 12$\,pc for HD\,23642.

\begin{table} \caption{ Identifications and astrophysical data for HD\,23642. 
\newline {\bf References:} (1) Perryman et al. (\cite{perryman}); (2) Abt \& Levato (\cite{abtlevato}); (3) 2 Micron All Sky Survey; (4) M04; quantities in parentheses are uncertainties in the final digit of the quanities.}
\label{table:photdata} \centering
\begin{tabular}{lr@{}lc} \hline \hline
                              &     & HD\,23642             & References\\\hline
Hipparcos number              &     & HIP 17704             & 1         \\
Hipparcos distance (pc)       &     & $111 \pm 12$          & 1         \\
Spectral type                 &     & A0\,Vp\,(Si) + Am     & 2         \\[2pt] 
$B_T$                         &     & 6.923 $\pm$ 0.011     & 1         \\
$V_T$                         &     & 6.839 $\pm$ 0.011     & 1         \\
$J_{\rm 2MASS}$               &     & 6.635 $\pm$ 0.023     & 3         \\
$H_{\rm 2MASS}$               &     & 6.641 $\pm$ 0.026     & 3         \\
$K_{\rm 2MASS}$               &     & 6.607 $\pm$ 0.024     & 3         \\[2pt] 
Orbital period (days)         &     & 2.46113400(34)        & 4         \\
Reference time (HJD)          &     & 2\,452\,903.5981(13)  & 4         \\
\hline \end{tabular} \end{table}


\section{Spectroscopic analysis}                                                                            \label{sec:spec}    

M04 observed HD\,23642 five times with the \'Elodie \'echelle spectrograph on the 1.93\,m telescope of the Observatoire de  Haute-Provence. The radial velocities derived were combined with the spectroscopic observations of Pearce (\cite{pearce}) and Abt (\cite{abt}), using lower weights for the older data, to calculate a circular spectroscopic orbit. 

\begin{figure} 
\resizebox{\hsize}{!}{\includegraphics{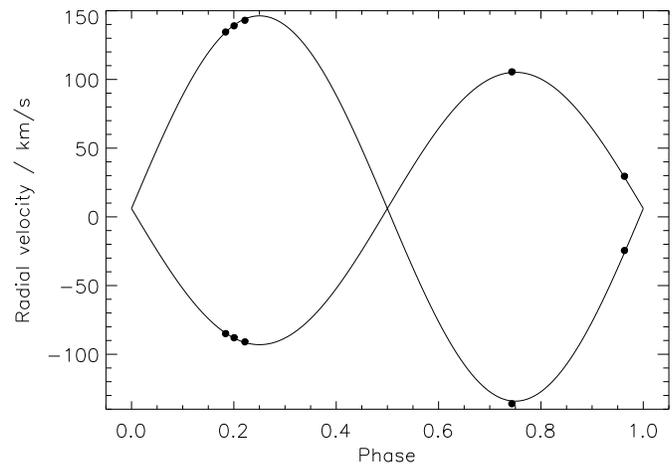}}
\caption{Spectroscopic orbit for HD\,23642}
\label{fig:specorbit}
\end{figure}

The low weight -- and low precison -- of the data of Pearce (\cite{pearce}) and Abt (\cite{abt}) mean that they contribute little to the accuracy of the spectroscopic orbit. For comparison with the results of M04 we have chosen to derive the orbit using only the five \'echelle velocities for each star. The orbit was computed using {\sc sbop}\footnote{Spectroscopic Binary Orbit Program written by Dr.\ P.\ B.\ Etzel (\texttt{http://mintaka.sdsu.edu/faculty/etzel/}).}, with the orbital ephemeris from M04, eccentricity fixed at zero, and equal systemic velocities for both stars. The root-mean-squares of the residuals of the resulting spectroscopic orbit are 0.4 and 1.2\kms\ for the primary and secondary stars, respectively. The spectroscopic orbit is plotted in Fig.~\ref{fig:specorbit} and its parameters are given in Table~\ref{table:specorbit}. The orbital parameters are in acceptable agreement with those of M04 and Torres (\cite{torres}).
 
\begin{table} \caption{Spectroscopic orbital parameters for HD\,23642.}
\label{table:specorbit} \centering
\begin{tabular}{l r@{\,$\pm$\,}l r@{\,$\pm$\,}l} \hline \hline 
                                          & \multicolumn{2}{c}{Primary} & \multicolumn{2}{c}{Secondary} \\ \hline
Semiamplitude $K$ (\kms)                  & 99.10 & 0.58 & 140.20 & 0.57 \\
Systemic velocity (\kms)                  & \multicolumn{4}{c}{6.07 $\pm$ 0.39} \\
Mass ratio $q$                            & \multicolumn{4}{c}{0.7068 $\pm$ 0.0050} \\
$a \sin i$ (\Rsun)                        & \multicolumn{4}{c}{11.636 $\pm$ 0.040} \\
$M \sin^3 i$ (\Msun)                      & 2.047 & 0.021 & 1.447 & 0.017 \\ \hline 
\end{tabular} \end{table}

\subsection{Determination of effective temperatures}        \label{specsynth}

\begin{figure*} \resizebox{\hsize}{!}{\includegraphics{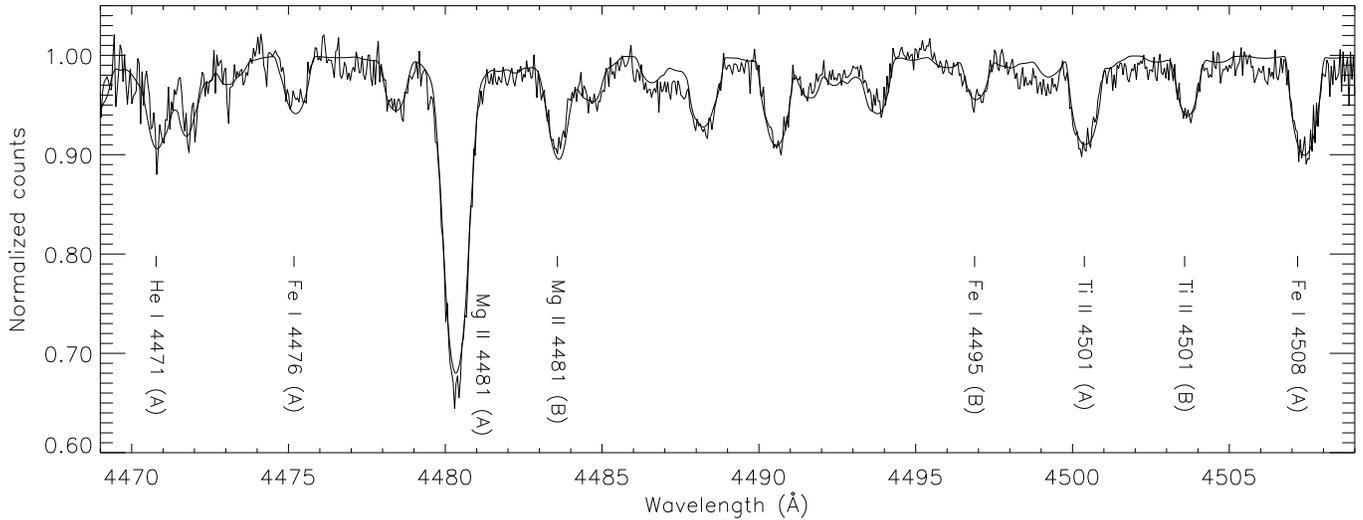}}
\caption{Comparison between a spectrum of HD\,23642 and the best-fitting synthetic spectrum used to determine the atmospheric parameters of the stars. The spectrum of HD\,23642 plotted here is a recombination of the individual spectra of the two stars, which were obtained by spectral disentangling (Simon \& Sturm \cite{SS}). The source ion of some lines of interest have been indicated, with the rest wavelength of the line (\AA) and which star is producing it (A for the primary or B for the secondary). The effective radial velocities of the primary and secondary stars in this diagram are $-58$ and $+160$\kms, respectively.}
\label{fig:specfit} \end{figure*}

Atmospheric parameters were derived for the components of HD\,23642 by comparing the observed spectra (from M04) with synthetic spectra calculated using {\sc uclsyn} (Smith \cite{smith}, Smalley et al.\ \cite{SSD}), Kurucz (\cite{kurucz}) {\sc atlas9} model atmospheres without approximate convective overshooting (Castelli, Gratton \& Kurucz \cite{CGK}) and absorption lines from the Kurucz \& Bell (\cite{KB}) linelist. The spectra were rotationally broadened as necessary and instrumental broadening was applied with a FWHM of 0.11\,\AA\ to match the resolution of the observations. Surface gravities of 4.25 were assumed for both stars. 

For the primary star, spectroscopic fitting gives an effective temperature $T_{\rm eff,A} = 9750 \pm 250$\,K with a microturbulent velocity of $\zeta_{\rm T,A} = 2$\kms\ and a rotational velocity of $V_{\rm A} \sin i = 37 \pm 2$\kms. For the secondary star we find $T_{\rm eff,B} = 7600 \pm 400$\,K, $\zeta_{\rm T,B} = 4$\kms\ and $V_{\rm B} \sin i = 32 \pm 3$\kms. The values for microturbulence velocity are consistent with those typically found for stars of these effective temperatures (Smalley \cite{smalley04}). The quoted uncertainties are limits of high confidence and are somewhat larger than the formal fitting errors. A monochromatic light ratio of $0.25 \pm 0.05$ was obtained at 4480\,\AA. The observations and the best-fitting synthetic spectrum are represented in Fig.~\ref{fig:specfit}.

Atmospheric parameters can be also estimated from photometric indices. We have obtained $uvby\beta$ photometry from Hauck \& Mermilliod (\cite{HM}) and dereddened it using $\Eby = 0.008$, calculated from $\EBV = 0.012$ (M04) and $\Eby \approx 0.73 \EBV$ (Crawford \cite{crawford}). Using the semi-empirical grid calibrations of Moon \& Dworetsky (\cite{MD}) and the {\sc tefflogg} program (Moon \cite{moon}), we obtained $T_{\rm eff} = 9200$ and $\log g = 4.30$, for the combined light of the system. To evaluate the effects of the secondary we have subtracted the photometry of the classical Am star 63\,Tauri using a $V$-filter magnitude difference of 1.44. This gave the parameters $T_{\rm eff,A} = 9870$\,K and $\log g_{\rm A} = 4.37$ for the primary component, in good agreement with our observationally determined parameters for this star.


A near-fundamental determination of effective temperature can be obtained using the Infrared Flux Method (Blackwell \& Shallis \cite{BS}). Ultraviolet fluxes were obtained from the IUE archive, optical fluxes from Kharitonov et al.\ (\cite{KTK}) and infrared fluxes from the 2MASS catalogue. From this the total integrated flux at the Earth was found to be $(5.44 \pm 0.44) \times 10^{-8}$ erg\,s$^{-1}$\,cm$^{-2}$, for a reddening of $\EBV = 0.012$ (taken from M04). The IRFM then yielded $T_{\rm eff} = 8900 \pm 350$\,K, which is rather low compared to the above values but is affected by the flux contribution of the cooler secondary star, which is proportionately brighter in the infrared. Allowing for the presence of the secondary star using the method of Smalley (\cite{smalley93}) we find that the primary would have $T_{\rm eff,A} = 9250 \pm 400$\,K for a secondary star with $T_{\rm eff,B} = 7500 \pm 500$\,K, which is consistent with the values determined above.

Fundamental effective temperatures can be obtained for binary systems using total integrated fluxes and angular diameters obtained from system parameters, and known distances (Smalley \& Dworetsky \cite{SD}; Smalley et al.\ \cite{SGKB}). In the case of HD\,23642 the properties of the system have been found using a model-dependent method, so application of this procedure would lead to a circular argument. However, the method does allow for a consistency check on the two effective temperatures and, importantly, their error estimates. Using the parameters obtained in the present work, we find $T_{\rm eff,A} = 9620 \pm 280$\,K and $T_{\rm eff,B} = 7510 \pm 430$\,K for the primary and secondary, respectively. Similar results are obtained for the parameters given by M04. However, use of the Hipparcos parallax of HD\,23642 (which gives a distance of $111 \pm 12$\,pc) would give $T_{\rm eff,A} = 8640 \pm 540$\,K and $T_{\rm eff,B} = 6690 \pm 570$\,K, which are clearly inconsistent with the values obtained above. The `short' Pleiades distance ($120 \pm 3$\,pc) would give $T_{\rm eff,A} = 9000 \pm 310$\,K and $T_{\rm eff,B} = 6970 \pm 450$\,K, which is closer but still somewhat discrepant.

Using several techniques we have found the effective temperatures of the two components stars of HD\,23642 to be $T_{\rm eff,A} = 9750 \pm 250$\,K for the primary and $T_{\rm eff,B} = 7600 \pm 400$\,K for the secondary. Our error estimates are higher than those reported in M04, primarily because we have assessed the influence of external uncertainties, in addition to the internal precision of fits to spectra.


\section{Photometric analysis}                                                                              \label{sec:photo}    

The $B$ and $V$ light curves contain 432 and 492 individual measurements, respectively, obtained with a 28\,cm Schmidt-Cassegrain telescope and photometer by M04. The two light curves were solved separately using {\sc ebop} (Nelson \& Davis \cite{ND}, Popper \& Etzel \cite{PE}). This is a simple and efficient light curve fitting code in which the discs of the stars are modelled using biaxial ellipsoids. Linear limb darkening coefficient values of 0.496 and 0.596 ($B$) and 0.421 and 0.548 ($V$), for the primary and secondary stars respectively, were adopted from van Hamme (\cite{hamme}) as the light curves are not of sufficient quality to include them as free parameters. Gravity darkening exponents $\beta_1$ were fixed at 1.0 (Claret \cite{claret98}) and the mass ratio was fixed at the spectroscopic value. The ephemeris given in M04 was used and the orbit was assumed to be circular. 

Although the M04 light curves are of reasonable quality, deriving accurate parameters from them is problematic due to the shallow eclipses. This is exacerbated by some scattered data in the $B$ light curve, which makes it less reliable than the $V$ light curve. As contaminating `third' light, $L_3$, is poorly constrained by the observations, we have made separate solutions for $L_3 = 0$ and 0.05 (in units of the total light of the eclipsing stars) and included differences in the parameter values derived in the uncertainties quote below. As there are no features in the spectra of HD\,23642 known to come from a third star, it is unlikely that third light is greater than 5\%.

Initial solutions provided an inadequate fit to the light variation outside eclipse so the reflection effect for the secondary star was separately adjusted towards best fit rather than being calculated from the system geometry. We also solved the light curves using the Wilson-Devinney code (Wilson \& Devinney \cite{WD}, Wilson \cite{wilson}), using the 1998 version ({\sc wd98}) with a detailed treatment of reflection. As the differences between the {\sc ebop} and {\sc wd98} solutions were negligible, further analysis was undertaken using {\sc ebop}. This code has two important advantages; a detailed error analysis is not prohibitively expensive in terms of computer time, and the philosophy of the {\sc ebop} code is to solve for the set of parameters most directly related to the light curve shape.

As the photometry of M04 is not supplied with observational errors, we have weighted all observations equally. We will judge the quality of the fit of an {\sc ebop} model light curve to the observational data using the root mean square of the residuals of the fit, $\sigma_{\rm rms}$.

\subsection{Light curve solution}

\begin{figure*} \resizebox{\hsize}{!}{\includegraphics{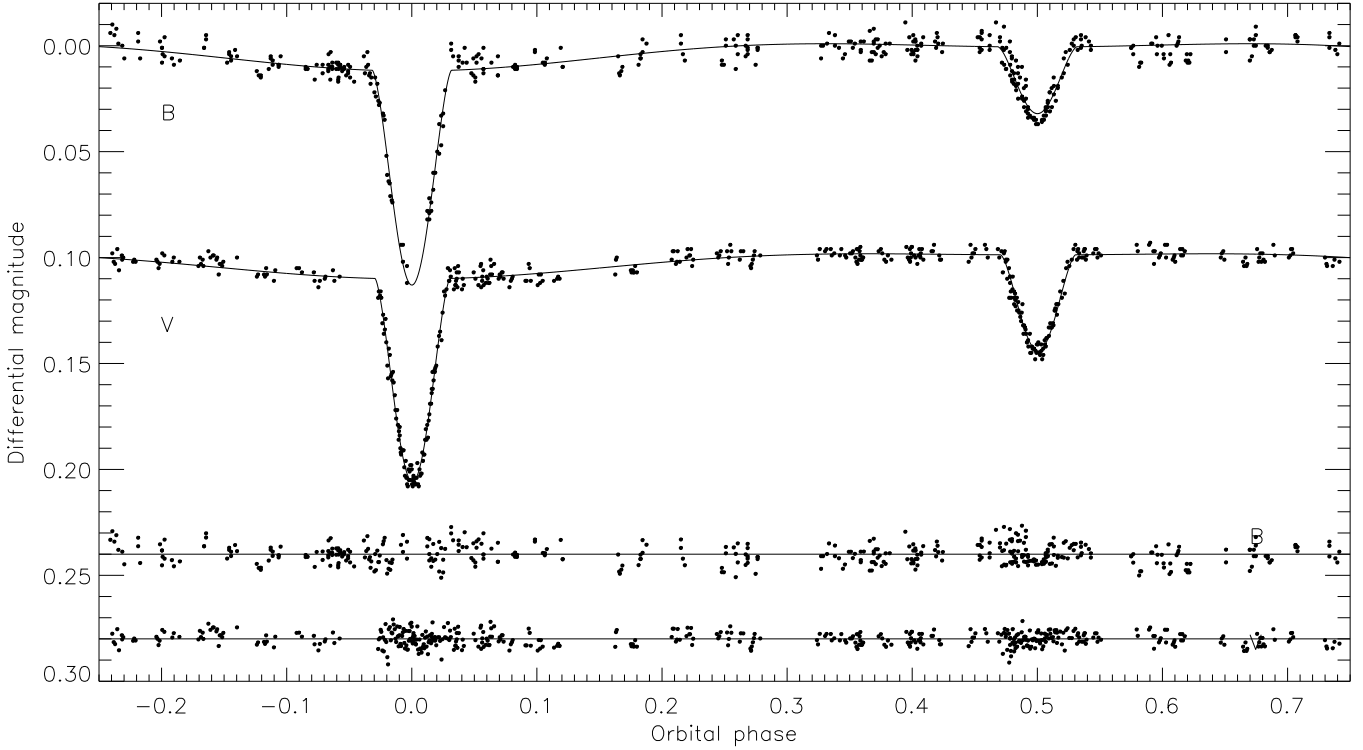}}
\caption{The M04 $B$ and $V$ light curves with our best fitted overplotted. The $V$ light curve is shifted by +0.1\,mag for clarity. The residuals of the fit are offset by +0.24\,mag and +0.28\,mag for the $B$ and $V$ light curves respectively. Note that the poor fit around the secondary eclipse in the $B$ light curve is due to scattered data, as suggested by the distribution of the residuals for this light curve. Rejection of the offending data makes the fit look better but is otherwise unjustified (see text for discussion).}
\label{fig:lcplot} \end{figure*}

\begin{table*} \caption{Photometric solution A, found by fitting the light curves without using a spectroscopic light ratio to constrain the ratio of the radii. The individual errors are from Monte Carlo simulations. The combined parameters (bottom line) are the weighted means of the two different values for each parameter. The quality of the fit is given by $\sigma_{\rm rms}$ (see text for definition).}
\label{table:lcsolution1} \centering
\begin{tabular}{l r@{\,$\pm$\,}l r@{\,$\pm$\,}l r@{\,$\pm$\,}l r@{\,$\pm$\,}l r@{\,$\pm$\,}l r@{\,$\pm$\,}l c} \hline \hline
Light & \mc{Surface bright-} & \mc{Light} & \mc{Primary radius,}   & \mc{Ratio of the} & \mc{Secondary radius,} & \mc{Orbital} & $\sigma_{\rm rms}$ \\
curve & \mc{ness ratio, $J$} & \mc{ratio} & \mc{$r_{\rm A}$ ($a$)} & \mc{radii, $k$}   & \mc{$r_{\rm B}$ ($a$)} & \mc{inclination, $i$ (\degr)} & ($m$mag) \\ \hline
$B$      &  0.316 & 0.018  &  0.264 & 0.083  &  0.1536 & 0.0069  &  0.929 & 0.116  &  0.1427 & 0.0108  &  76.91 & 0.41 & 4.639 \\
$V$      &  0.493 & 0.018  &  0.366 & 0.094  &  0.1487 & 0.0057  &  0.882 & 0.093  &  0.1312 & 0.0083  &  77.80 & 0.24 & 3.254 \\
Combined &      \mc{}      &      \mc{}      &  0.1507 & 0.0044  &  0.900 & 0.073  &  0.1355 & 0.0066  &  77.57 & 0.21 &   \\ \hline
\end{tabular} \end{table*}

\begin{table*} \caption{Photometric solution B, found using {\sc ebop} and the light ratio of Torres (\cite{torres}). The individual uncertainties are from Monte Carlo simulations with a fixed ratio of the radii, $k$. Symbols have the same meaning as in Table~\ref{table:lcsolution1}. The adopted parameter values (bottom line) are the weighted means of the different parameter values for each light curve. The adopted uncertainties depend mainly on the possible values of $k$.} 
\label{table:lcsolution2} \centering
\begin{tabular}{l r@{\,$\pm$\,}l r@{\,$\pm$\,}l  r@{\,$\pm$\,}l r@{\,$\pm$\,}l r@{\,$\pm$\,}l r@{\,$\pm$\,}l c} \hline \hline 
Light   & \mc{Light}    & \mc{}       & \mc{}         & \mc{$r_{\rm A}$}& \mc{$r_{\rm B}$}& \mc{$i$}     & $\sigma_{\rm rms}$  \\
curve   & \mc{ratio}    & \mc{$k$}    & \mc{$J$}      & \mc{($a$)}      & \mc{($a$)}      & \mc{(\degr)} & ($m$mag)     \\ \hline
$V$     & \mc{0.300}    & \mc{0.8075} & 0.483 & 0.014 & 0.1531 & 0.0009 & 0.1237 & 0.0007 & 78.04 & 0.08 & 3.254        \\
$V$     & \mc{0.335}    & \mc{0.8477} & 0.489 & 0.015 & 0.1508 & 0.0009 & 0.1278 & 0.0008 & 77.90 & 0.08 & 3.254        \\
$V$     & \mc{0.370}    & \mc{0.8858} & 0.493 & 0.015 & 0.1485 & 0.0009 & 0.1315 & 0.0008 & 77.79 & 0.08 & 3.254        \\[2pt]
$B$     & 0.190 & 0.008 & \mc{0.8075} & 0.304 & 0.014 & 0.1599 & 0.0019 & 0.1291 & 0.0015 & 77.47 & 0.19 & 4.654        \\
$B$     & 0.214 & 0.008 & \mc{0.8477} & 0.309 & 0.015 & 0.1580 & 0.0019 & 0.1339 & 0.0016 & 77.26 & 0.19 & 4.647        \\
$B$     & 0.237 & 0.009 & \mc{0.8858} & 0.313 & 0.015 & 0.1560 & 0.0019 & 0.1382 & 0.0016 & 77.08 & 0.19 & 4.642        \\[2pt]
Adopted &    \mc{ }     & 0.848&0.039 &    \mc{ }     & 0.1538 & 0.0024 & 0.1300 & 0.0037 & 77.78 & 0.17 &              \\ \hline
\end{tabular} \end{table*}

The ratio of the radii, $k$, is poorly constrained by the light curves because the eclipses are very shallow. Whilst a reasonable photometric solution can be obtained from the light curves alone, an alternative is to use a spectroscopic light ratio to constrain $k$. A light ratio of $\frac{l_{\rm B}}{l_{\rm A}} = 0.31 \pm 0.03$ was given by Torres (\cite{torres}), based on a cross-correlation analysis of a 45\,\AA\ wide spectral window centred on 5187\,\AA. This spectroscopic light ratio is noted to be preliminary so we provide separate solutions without (`solution A') and with (`solution B') its inclusion in the light curve fitting procedure, but preference is given to solutions including the spectroscopic constraint. We have used the light ratio of Torres (\cite{torres}) as it is based on a larger amount of observational data than the light ratio found in Section~\ref{specsynth}, and because the wavelength it was obtained at is closer to the central wavelength of the $V$ filter.

The light ratio of Torres (\cite{torres}) was converted to a $V$ filter light ratio using a $V$ filter response function and synthetic spectra, calculated from {\sc atlas9} model atmospheres, for the effective temperatures and surface gravities found in our preliminary analyses. The resulting $V$ filter light ratio of $0.335 \pm 0.035$ (where the uncertainties include a small contribution due to possible systematic errors from the use of {\sc atlas9} model atmospheres) has been used to constrain $k$ using the $V$ light curve. The resulting values of $k$ were then adopted for solution of the $B$ light curve. 

Table~\ref{table:lcsolution1} gives solution A, and Table~\ref{table:lcsolution2} gives solution B. The best fit for the former solution is compared to the observational data in Fig.~\ref{fig:lcplot}; the light variation for the latter solution is almost identical so has not been plotted. Note that the $B$ light curve appears to be badly fitted at the centre of the secondary eclipse. Investigation has revealed that this is not a problem with the {\sc ebop} model, but is caused by scatter present in the observational data. A better fit can be obtained by rejecting one night's data, around phase 0.48, which is brighter than the model light curve. After rejection of these data, the fit is significantly improved but the derived parameters are quite similar. We have therefore included all observational data in the fitting procedure, and suggest that further photometric data should be obtained.

\begin{figure} \resizebox{\hsize}{!}{\includegraphics{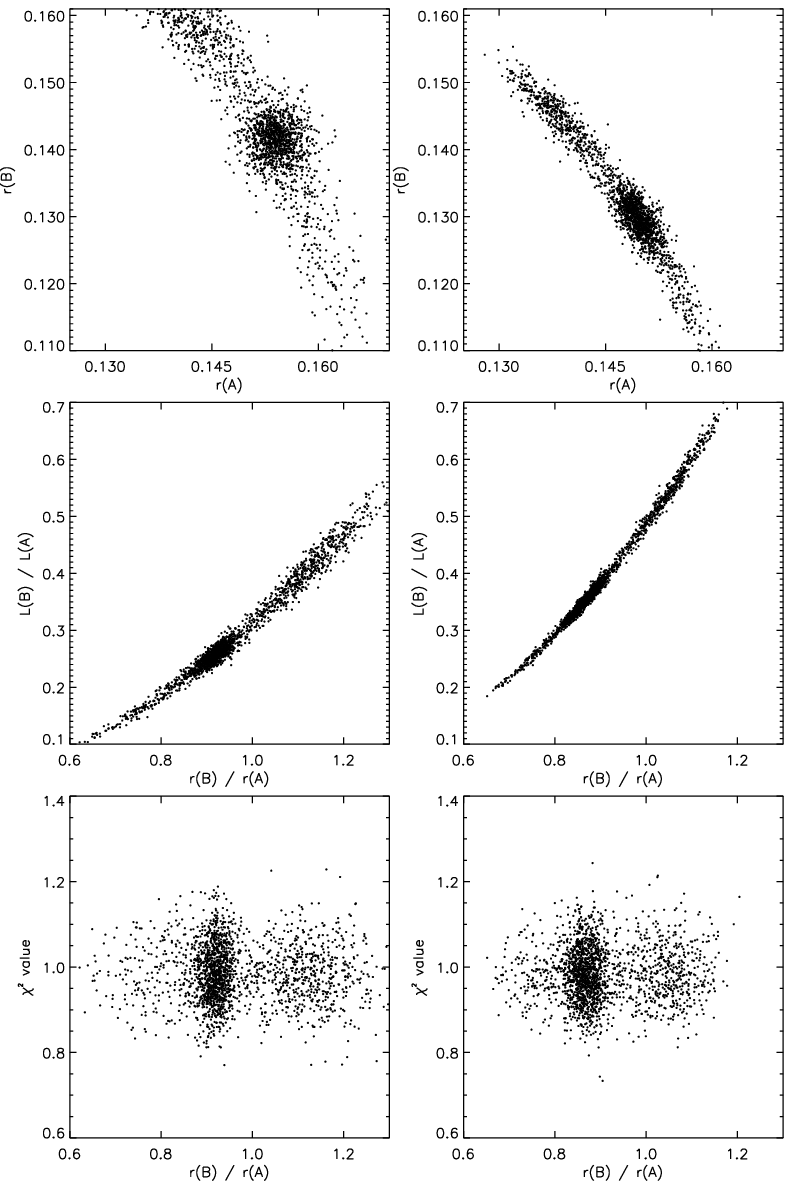}}
\caption{Results of the Monte Carlo analysis for the $B$ (left panels) and $V$ (right panels) light curves. The top two panels show $r_{\rm A}$ plotted against $r_{\rm B}$. The middle two panels show $k$ versus the light ratios. The lower two panels show the reduced $\chi^2$ versus $k$. These values were calculated using the residuals of the best fits to the observed light curves. Note the greater scatter of the Monte Carlo solutions for the $B$ light curve, which is due to the larger observational errors. Only 2000 of the 10\,000 points have been plotted in each panel.} 
\label{fig:lcerrs}\end{figure}

The uncertainties in the fitted parameters were estimated using Monte Carlo simulations (see Southworth et al. \cite{SMSb} for details). In this algorithm, a best fit is found and the observational scatter is calculated. The model is then evaluated at the phases of observation, random Gaussian noise (of the same size as the actual observational scatter) is added, and the resulting synthetic light curve is fitted. This procedure is undertaken 10\,000 times for each observed light curve. The uncertainties in the photometric parameter values are estimated by calculating the standard deviation of the values found during the Monte Carlo simulations. The resulting uncertainties are included in Table~\ref{table:lcsolution1} and Table~\ref{table:lcsolution2}. Some results of the Monte Carlo simulations are shown in Fig.~\ref{fig:lcerrs}. Uncertainties in the theoretically-derived limb darkening coefficients have been incorporated by perturbing the values of the coefficients by $\pm$0.05, on a flat distribution, for each Monte Carlo simulation. The uncertainties estimated using this Monte Carlo simulation algorithm have been found by the authors to be extremely reliable (Southworth et al. \cite{SMSc}, \cite{SMSb}).

\begin{figure} \resizebox{\hsize}{!}{\includegraphics{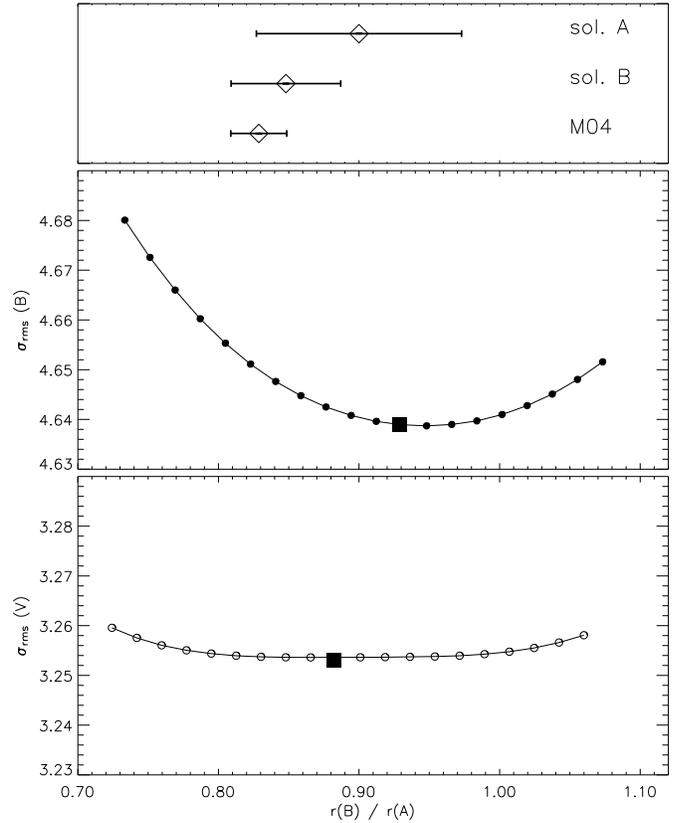}}
\caption{Comparison between the quality of the fit for different values of $k$ and the solutions derived in this work and by M04. $\sigma_{\rm rms}$ has been plotted against $k$ for the $B$ light curve (middle panel) and $V$ light curve (lower panel) using the same scales. The optimum values of $k$ found in the individual solutions of the two light curves are indicated by filled squares. The upper panel shows the values and uncertainties of $k$ found in solution A, solution B, and by M04. Note that the quality of the photometry is excellent, particularly for the $V$ light curve where the photometric error of an individual datapoint is 3.25 millimagnitudes.} 
\label{fig:lcresidk}\end{figure}

The fractional stellar radii given by solution A are $r_{\rm A} = 0.1507 \pm 0.0044$ and $r_{\rm B} = 0.1355 \pm 0.0066$, whereas inclusion of the spectroscopic light ratio (solution B) gives $r_{\rm A} = 0.1538 \pm 0.0024$ and $r_{\rm B} = 0.1300 \pm 0.0037$. The fractional radii found by M04 were $r_{\rm A} = 0.1514 \pm 0.0025$ and $r_{\rm B} = 0.1254 \pm 0.0022$. Our spectroscopic-constraint result agrees well with the M04 result, but we are unable to reproduce the small uncertainties claimed by M04. Fig.~\ref{fig:lcresidk} compares the values of $k$ found for solution A, solution B and by M04, to the residuals of the fit for a range of different values of $k$ for each light curve. Solution A has the largest uncertainty but is close to the minima of the residuals in the $B$ and $V$ light curves. Solution B, for which $k$ was found using a spectroscopic light ratio, has a smaller uncertainty in $k$ but has a slight dependence on theoretical model atmospheres. The results of M04 were obtained from the same data as our solution A, but the two values of $k$ are quited different. The low uncertainties quoted by M04 mean that their value of $k$ (calculated by us from the stellar radii given by M04) is inconsistent with the minima in the residuals curves. M04 adopted formal errors from their photometric analysis, which are known to be generally somewhat optimistic (Popper \cite{popper}).


\section{Absolute dimensions and comparison with stellar models}

\begin{table} \caption{Absolute dimensions of the components of HD\,23642, calculated using photometric solution B. Symbols have their usual meanings and the equatorial rotational velocities are denoted by $V_{\rm eq}$. The absolute bolometric magnitudes have been calculated using a solar luminosity of 3.826$\times$10$^{26}$\,W and absolute bolometric magnitude of 4.75.}
\label{table:dimensions} \centering
\begin{tabular}{l c r@{\,$\pm$\,}l c r@{\,$\pm$\,}l} \hline \hline
              & \hspace{10pt} & \multicolumn{2}{c}{Primary star} & \hspace{5pt} & \multicolumn{2}{c}{Secondary star} \\ \hline
Mass (\Msun)                  &  &    2.193 &   0.022 &  &    1.550 &   0.018  \\
Radius (\Rsun)                &  &    1.831 &   0.029 &  &    1.548 &   0.044  \\
$\log g$ (cm\,s$^{-1}$)       &  &    4.254 &   0.014 &  &    4.249 &   0.025  \\
\Teff\ (K)                    &  &  9750    &  250    &  &  7600    &  400     \\
$\log (L/\lsun)$              &  &    1.437 &   0.047 &  &    0.858 &   0.095  \\
\Mbol                         &  &    1.16  &   0.12  &  &    2.60  &   0.24   \\
$V_{\rm eq}$ (\kms)           &  &   38     &   2     &  &   33     &   3      \\
\Vsync\ (\kms)                &  &   37.6   &   0.6   &  &   31.8   &   0.9    \\ \hline
\end{tabular} \end{table}

The absolute dimensions of the two stars have been calculated from the spectroscopic results and photometric solution B, and are given in Table~\ref{table:dimensions}. We note that the radii of the two stars calculated using solution A are $R_{\rm A} = 1.796 \pm 0.053$\Rsun\ and $R_{\rm B} = 1.615 \pm 0.079$\Rsun. The masses and radii of the two components of HD\,23642 can be compared to stellar evolutionary models to estimate the metal abundance of the system. This is important because unusual chemical compositions have been suggested as possible reasons why the `long' distance scale of the Pleiades is in disagreement with the Hipparcos parallax distances.

\begin{figure} 
\resizebox{\hsize}{!}{\includegraphics{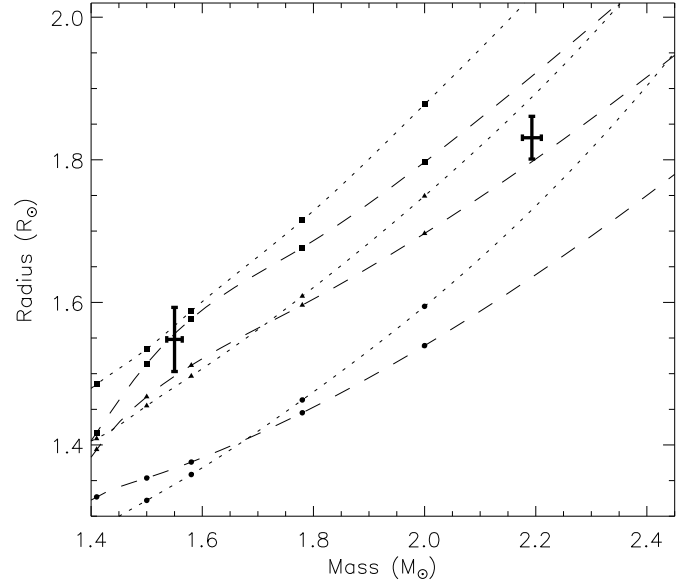}}
\caption{Comparison between the observed properties of HD\,23642 and the Granada stellar evolutionary models for metal abundances of $Z=0.01$ (circles), $Z=0.02$ (triangles) and $Z=0.03$ (squares). Predictions for normal helium abundances are plotted with dashed lines and helium-rich model predictions are plotted using dotted lines. An age of 125\,Myr was assumed.} 
\label{fig:modelfit1} \end{figure}

\begin{figure} 
\resizebox{\hsize}{!}{\includegraphics{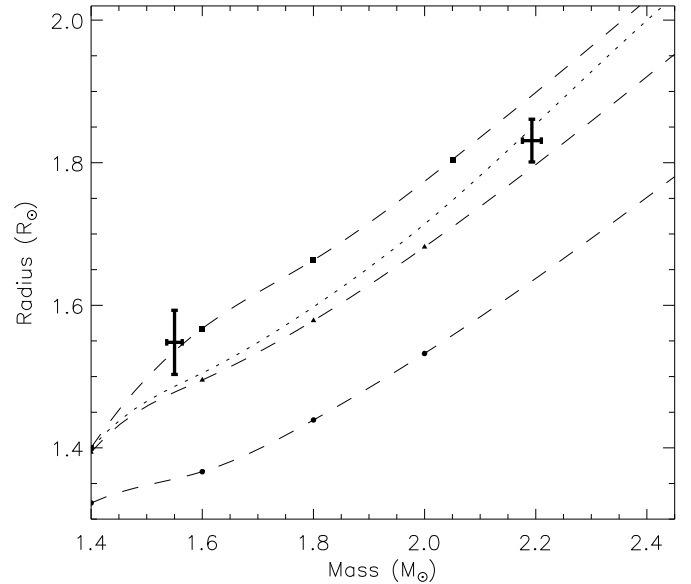}}
\caption{Comparison between the observed properties of HD\,23642 and the Cambridge stellar evolutionary models for metal abundances of $Z=0.01$ (circles), $Z=0.02$ (triangles) and $Z=0.03$ (squares). An age of 125\,Myr was assumed (dashed lines), but predictions for a solar chemical composition and an age of 175\,Myr are also shown (dotted line).} 
\label{fig:modelfit2} \end{figure}

In Fig.~\ref{fig:modelfit1} the masses and radii of the components of HD\,23642, found from solution B, have been compared to predictions of the Granada stellar evolutionary models (Claret \cite{claret95}, \cite{claret97}; Claret \& Gim\'enez \cite{CG}).
An age of 125\,Myr has been adopted (Stauffer, Schultz \& Kirkpatrick \cite{SSK}); a change in this age by $-25$ or $+75$\,Myr does not affect the conclusions below. Fig.~\ref{fig:modelfit1} shows predictions for metal abundances of $Z = 0.01$, 0.02 and 0.03. For each metal abundance we have plotted predictions for normal helium abundance (dashed lines) and for significantly enhanced helium abundances (dotted lines). Fig.~\ref{fig:modelfit2} compares the masses and radii of HD\,23642 to the predictions of the Cambridge stellar evolutionary models (Pols et al. \cite{pols}) for metal abundances of $Z = 0.01$, 0.02 and 0.03. The solar chemical composition isochrone is also shown for an age of 175\,Myr.

The masses and radii of the components of HD\,23642 suggest that the metal abundance of the dEB is the slightly greater than solar ($Z \approx 0.02$). Predictions for enhanced helium abundances are ruled out as they predict a mass-radius relation for HD\,23642 much steeper than observed. The approximately solar Pleiades iron abundance found by Boesgaard \& Friel (\cite{BF}), from high-resolution spectroscopy of F dwarfs, is confirmed. The `short' and `long' Pleiades distances therefore cannot be reconciled by adopting an unusual chemical composition for the cluster.

\section{The distance to HD\,23642 and the Pleiades}                                                         \label{sec:distance}

We will now derive the distance to HD\,23642 using three methods, one of which is introduced here. We investigate the normal technique using bolometric corrections to find absolute visual magnitudes, the use of surface brightness calibrations in terms of observed colour indices, and a new method based on infrared surface brightness calibrations in terms of effective temperature. All the techniques require reliable apparent magnitudes, which for HD\,23642 are available from the Tycho experiment on the Hipparcos satellite and from 2MASS. The apparent $B$ and $V$ magnitudes observed by Tycho (Perryman et al. \cite{perryman}) have been converted to the Johnson photometric system using the calibration of Bessell (\cite{bessell}). The $JHK$ photometry from 2MASS has been converted to the SAAO near-infrared photometric system using the calibration of Carpenter (\cite{carpenter}).

\subsection{Distance from the use of bolometric corrections}

The traditional method of determining the distance of an eclipsing binary is to calculate the luminosity of each star from its radius and effective temperature. The resulting values of absolute bolometric magnitude, \Mbol, are then converted to absolute visual magnitudes, $M_V$, using bolometric corrections (BCs). The combined $M_V$ of the two stars is then compared to the apparent visual magnitude, $m_V$, to find the distance modulus. This method is quite capable of yielding accurate results provided that care is taken to use effective temperature measurements consistent with the fundamental definition of effective temperature. Consistent zeropoints must also be used for the calculation of absolute bolometric magnitude and the BCs (Bessell, Castelli \& Plez \cite{BCP}).

To find the distance to HD\,23642, we have adopted the astrophysical parameters of the system given in Table~\ref{table:dimensions}. An interstellar reddening of $\EBV = 0.012 \pm 0.004$\,mag has been adopted from M04. We have calculated the distance to HD\,23642 using the empirical BCs given by Code et al. (\cite{code}), with a calibration uncertainty of 0.05\,mag (Clausen 2004) added in quadrature. We have also derived the distance using the theoretically-calculated BCs of Bessell et al.\ (\cite{BCP}) and of Girardi et al.\ (\cite{girardi}). For the primary component of HD\,23642, the uncertainty resulting from its effective temperature measurement is reduced due to the form of the BC function around 10\,000\,K. The main contribution to the overall uncertainty comes from the uncertainties in the effective temperatures of the stars. 

The distances found by using BCs are given in Table~\ref{table:bcdistance} and show that a value around 139\,pc is obtained consistently. The distance found using the empirically derived BCs of Code et al. (\cite{code}) is very similar to the distances found by using the theoretically-derived BCs, suggesting that systematic errors due to the use of model atmospheres are small. Distances derived using BCs for the $JHK$ filters are more precise because the uncertainties in effective temperature and interstellar reddening are smaller. If the results of photometric solution A are used to find the distance to HD\,23642 using this method, distances of around 139\,pc are also found, but with slightly larger uncertainties. 

The BCs given by Girardi et al. (\cite{girardi}) are available for several different metal abundances, [M/H]. We have investigated the effect of non-solar metal abundances on the distance derived using BCs for ${\rm [M/H]} = -0.5$ and +0.5 (Table~\ref{table:bcdistance}). Whilst a significant change in distance is found for distances derived using BCs for the $V$ filter, the effect is much smaller for the $K$ filter, underlining the usefulness of infrared photometry in determining distances to dEBs.

\begin{table} \caption{The distances derived for HD\,23642 by using several different sources of bolometric corrections.}
\label{table:bcdistance} \centering
\begin{tabular}{lccr@{\,$\pm$\,}l} \hline \hline
Bolometric corrections              & Filter    & [M/H]     & \mc{Distance (pc)}    \\ \hline
Code et al. (\cite{code})           & $V$       &           &     138.1 & 6.2       \\
Bessell et al. (\cite{BCP})         & $V$       &           &     139.5 & 5.3       \\
Bessell et al. (\cite{BCP})         & $K$       &           &     138.6 & 3.3       \\
Girardi et al. (\cite{girardi})     & $B$       & 0.0       &     140.2 & 6.1       \\ 
Girardi et al. (\cite{girardi})     & $V$       & 0.0       &     139.8 & 5.3       \\ 
Girardi et al. (\cite{girardi})     & $J$       & 0.0       &     138.7 & 3.8       \\ 
Girardi et al. (\cite{girardi})     & $H$       & 0.0       &     138.0 & 3.3       \\ 
Girardi et al. (\cite{girardi})     & $K$       & 0.0       &     138.8 & 3.3       \\ \hline 
Girardi et al. (\cite{girardi})     & $V$       & $-0.5$    &     137.9 & 5.3       \\ 
Girardi et al. (\cite{girardi})     & $V$       & $+0.5$    &     142.4 & 5.3       \\
Girardi et al. (\cite{girardi})     & $K$       & $-0.5$    &     137.9 & 3.3       \\ 
Girardi et al. (\cite{girardi})     & $K$       & $+0.5$    &     139.8 & 3.3       \\ \hline
\end{tabular} \end{table}

M04 found the distance to HD\,23642, using the Bessell et al. (\cite{BCP}) BCs, to be $131.9 \pm 2.1$\,pc. However, we have been unable to reproduce their result using the properties of HD\,23642 found by these authors. The absolute bolometric magnitudes given by M04 appear to have been taken from output files produced by the Wilson-Devinney light curve modelling code, which uses a solar luminosity of $\lsun = \sci{3.906}{26}$\,W and absolute bolometric magnitude of $\Mbolsun = 4.77$. The BCs used by M04, however, were calculated using different values: $\Lsun = \sci{3.855}{26}$\,W and $\Mbolsun = 4.74$ (Bessell et al. \cite{BCP}). This inconsistency appears to be sufficient to explain the discrepancy between their result and our results. If we adopt the masses, radii and effective temperatures of M04, we find a distance of $135.5 \pm 2.3$\,pc. This is about 3\,pc smaller than most of our results, mainly due to the smaller stellar radii found by M04 for the components of HD\,23642.

The discussion above shows that it is possible to calculate consistent distances to eclipsing binaries, using different sets of BCs.
One weakness of this method is that it is difficult to evaluate the systematic error introduced by the uncertainty in the zeropoint of the effective temperature scale. The use of BCs derived from theoretical models also introduces a systematic error which is likely to be small in this case but, in general, is difficult to quantify. This systematic error is due to deficiencies in the model, e.g., the approximate treatment of convection and the lack of complete spectral line lists. It may also be the case that the star has properties which are not accounted for by the model, e.g., spectral peculiarity due to magnetic fields or slow rotation. For these reasons it is desirable to develop an empirical distance determination method which is less sensitive to systematic errors in effective temperature and in which the sources of uncertainty are more explicit.

\subsection{Distance from the use of relations between surface brightness and colour indices}

The concept of surface brightness was first used in the analysis of dEBs about one hundred years ago (Kruszewski \& Semeniuk \cite{KS}). Effective temperatures of the components of several dEBs were found from the stellar radii and distances determined using parallax measurements, and relations between surface brightness and spectral type were investigated. 

Barnes \& Evans (\cite{BE}; Barnes, Evans \& Parsons \cite{BEP}) provided a calibration between visual surface brightness, $S_V$, and the $V-R$ colour index. The Barnes-Evans relation was used by Lacy (\cite{lacy}) to find the distances to nine dEBs. Popper (\cite{popper80}) presented calibrations between $S_V$ and the $B-V$ and $b-y$ photometric indices. Surface brightness calibrations for broad-band filter indices have recently been presented by Fouqu\'e \& Gieren (\cite{FG}), Di Benedetto (\cite{benedetto}),  Beuermann et al.\ (\cite{beuermann}) and Kervella et al.\ (\cite{KTDS}, hereafter KTDS04). Salaris \& Groenewegen (\cite{SG}) have calibrated the relation between $S_V$ and the Str\"omgren photometric index, $c_1$, for determination of the distances to B stars in the Magellanic Clouds.

In this method the angular diameter of the star is calculated and compared to its linear diameter, determined from photometric analysis, to find the distance. The advantage of this procedure is that the calculations can be entirely empirical, depending on the way in which the surface brightness calibration is constructed. One disadvantage is that individual apparent magnitudes and colour indices of the components of the dEB, calculated from the light ratios found in light curve analyses, must be used. The uncertainties in these quantities can cause the derived distance to have a low precision.

We have determined a distance to HD\,23642 of $138 \pm 19$\,pc using the $S_V$ versus $B-V$ calibration given by Di Benedetto (\cite{benedetto}). The main uncertainty in this result comes from the uncertainties in the light ratios in the $B$ and $V$ filters. The most recent calibration, given by KTDS04, has not been used because it does not allow for the nonlinear dependence of $S_V$ on $B-V$. The $B-V$ colour index is also not a good indicator of surface brightness because the $B$ filter is known to be sensitive to metallicity through the effect of line blanketing. The $V-K$ and $B-L$ indices are good surface brightness indicators (KTDS04; Di Benedetto \cite{benedetto}) because they are more sensitive to surface brightness and because the intrinsic scatter in the calibrations falls below 1\%.

\subsection{Distance from the use of relations between surface brightness and effective temperature}

Empirical relations between effective temperature and the wavelength-dependent surface brightness of a star, $S_{m_\lambda}$ where $m_\lambda$ is an apparent magnitude in filter $\lambda$, have been derived by KTDS04 from interferometric observations. This allows the surface brightnesses of the components of a dEB to be found without using the light ratios of the system found during the light curve analysis.

The zeroth-magnitude angular diameter is a measure of surface brightness and is defined to be the angular diameter a star would have if $m_\lambda = 0$: 
\begin{equation} \label{eq:zmld}
\phi^{(m_\lambda = 0)} = \phi \cdot 10^\frac{m_\lambda}{5} = \frac{S_{m_\lambda}}{5}
\end{equation}
where $\phi$ is the actual angular diameter of a star (van Belle \cite{belle}). From consideration of the definitions of surface brightness and angular diameter, it can be shown that the distance, $d$, to a dEB is given by
\begin{equation} \label{eq:d}
d = 10^{0.2 m_\lambda} \sqrt{\left[\frac{2 R_{\rm A}}{\phi_{\rm A}^{(m_\lambda = 0)}}\right]^2 + 
                             \left[\frac{2 R_{\rm B}}{\phi_{\rm B}^{(m_\lambda = 0)}}\right]^2 }
\end{equation}
where the stellar radii $R_{\rm A}$ and $R_{\rm B}$ are in AU, $\phi_{\rm B}^{(m_\lambda = 0)}$ are in arcseconds and distance is in parsecs. 

Calibrations for $\phi^{(m_\lambda = 0)}$ are given in terms of effective temperature by KTDS (where they are denoted using ZMLD$_\lambda$). These calibrations are for the broad-band $UBVRIJHKL$ filters and are valid for effective temperatures between 10\,000\,K and 3600\,K for main sequence stars. The $JHKL$ filter calibrations have the least scatter and are also the least affected by interstellar reddening. For the $K$ filter the scatter around the calibration is undetectable at a level of 1\% so we conservatively adopt a scatter of 1\%. We have applied the calibrations for $B$ and $V$ (derived from Tycho data) and for $J$, $H$ and $K$ (derived from 2MASS data). As there is no ``standard'' infrared photometric system, the calibrations of KTDS04 use data from several different $JHKL$ systems, so the systematic uncertainty of not having a standard system is already included in the quoted scatter in the calibrations. For the A stars only, the scatter around the $B$ and $V$ calibrations is much smaller than the overall scatter quoted, so for HD\,23642 the $B$ and $V$ distance uncertainties are overestimated by a factor of about two.

\begin{table} \caption{The results and individual error budgets for distance estimates using the effective temperatures and overall apparent magnitudes of the HD\,23642 system in the $BVJHK$ filters. All distances are given in parsecs and the total uncertainties are the sums of the individual uncertainties added in quadrature.}
\label{table:distance} \centering
\begin{tabular}{lrrrrr} \hline \hline
Uncertainty source            &      $B$ &      $V$ &      $J$ &      $H$ &      $K$ \\ \hline
Spectroscopic $K_{\rm A}$     &      0.3 &      0.3 &      0.3 &      0.3 &      0.3 \\
Spectroscopic $K_{\rm B}$     &      0.3 &      0.3 &      0.3 &      0.3 &      0.3 \\
Orbital inclination, $i$      &      0.1 &      0.1 &      0.1 &      0.1 &      0.1 \\
Fractional radius, $r_{\rm A}$&      1.8 &      1.7 &      1.5 &      1.4 &      1.4 \\
Fractional radius, $r_{\rm B}$&      0.8 &      1.0 &      1.3 &      1.4 &      1.5 \\
Primary \Teff                 &      4.8 &      3.3 &      1.7 &      0.7 &      0.7 \\
Secondary \Teff               &      3.8 &      3.5 &      2.1 &      1.5 &      1.4 \\
Reddening \EBV                &      1.1 &      0.8 &      0.3 &      0.2 &      0.1 \\
Apparent magnitude            &      1.0 &      1.0 &      1.9 &      1.9 &      1.9 \\
``Cosmic'' scatter            &      9.0 &      8.3 &      1.7 &      1.9 &      1.4 \\ \hline
Total uncertainty             &     11.1 &      9.9 &      4.2 &      3.8 &      3.5 \\
Distance                      &    142.8 &    141.4 &    139.6 &    138.4 &    139.1 \\ \hline 
\end{tabular} \end{table}

The distances found using equation~\ref{eq:d} and the KTDS04 calibrations are given in Table~\ref{table:distance}. We have calculated the distances from the results of the light curve solution and the spectroscopic velocity semiamplitudes ($K_{\rm A}$ and $K_{\rm B}$), in order to carefully assess how important the uncertainties in each quantity are to the final distance uncertainty. The intrinsic scatter in the calibrations is marked as ``cosmic'' scatter in Table~\ref{table:distance} and is the main contributor to the uncertainty in the $B$ and $V$ filter distances. Note that the uncertainties in the $JHK$ calibration distances are much smaller than in the $BV$ calibration distances, because the calibrations have much smaller scatter and reddening is less important. We will adopt the $K$ filter calibration distance of $139.1 \pm 3.5$\,pc as our final distance to HD\,23642 and so to the Pleiades. Note that we cannot treat any of the distance estimates investigated above as being independent of each other as they are all calculated using the same values for reddening, stellar radii and effective temperatures. If we adopt the results of photometric solution A, the $K$ filter distance we find is 139.7\,pc with an error of 4.7\,pc, which is entirely consistent with our adopted distance of $139.1 \pm 3.5$\,pc.

One shortcoming of finding distances using equation~\ref{eq:d} is that the effective temperature scales used in analysis of the dEB and for the calibration must be the same to avoid systematic errors. This is, however, a more relaxed constraint on the effective temperature scale than that involved in finding distance using BCs. We note that our effective temperature uncertainties include contributions due to possible spectral peculiarity and systematic offset relative to the (inhomogeneous) effective temperatures used in the KTDS04 calibration. The uncertainty in distance could be reduced by further observations and estimations of the effective temperatures of the two stars, using the same technique as for the stars used to calibrate the surface brightness relations.

                    
\section{Conclusion}                                                                                        \label{sec:conclusion}

\begin{table} \caption{Summary of the different distances found for the Pleiades or for Pleiades members, both from the literature (upper part of the table) and for HD\,23642 in this work (lower part of the table). References are also given in the text and abbreviations and symbols have their usual meanings.}
\label{table:summary} \centering
\begin{tabular}{l l@{\,$\pm$\,}l} \hline \hline
Source of distance measurement                                    & \mc{Distance (pc)}        \\ \hline
Hipparcos parallaxes (van Leeuwen \cite{leeuwen04})               & 120       & \ \ 3         \\  
Hipparcos parallax of HD\,23642                                   & 111       & 12            \\
Hipparcos parallax of HD\,23850                                   & 117       & 14            \\
MS fitting (Stello \& Nissen \cite{SN})                           & 132.4     & \ \ 1.8       \\
Ground-based parallaxes                                           & 130.9     & \ \ 7.4       \\
HST parallaxes of three Pleiades stars                            & 134.6     & \ \ 3.1       \\
Narayanan \& Gould (\cite{NG})                                    & 131       & 11            \\
Makarov (\cite{makarov})                                          & 129       & \ \ 3         \\
Eclipsing binary HD\,23642 (Munari et al.\ \cite{M04})            & 131.9     & \ \ 2.1       \\ 
Astrometric binary HD\,23850                                      & 132       & \ \ 4         \\ \hline
Code et al.\ (\cite{code}) empirical BCs ($V$ filter)             & 138.1     & \ \ 6.2       \\
Girardi et al.\ (\cite{girardi}) theoretical BCs ($V$ filter)     & 139.8     & \ \ 5.3       \\
Girardi et al.\ (\cite{girardi}) theoretical BCs ($K$ filter)     & 138.8     & \ \ 3.3       \\
Surface brightness--($B-V$) relation                              & 138       & 19            \\
Surface brightness--\Teff\ relation ($V$ filter)                  & 141.4     & \ \ 9.9       \\
Surface brightness--\Teff\ relation ($K$ filter)                  & 139.1     & \ \ 3.5       \\ \hline
\end{tabular} \end{table}

The `long' distance scale of the Pleiades is $132 \pm 3$\,pc and is supported by main sequence fitting, the distance of the astrometric binary Atlas, and by ground-based and Hubble Space Telescope parallax measurements. The `short' distance is $120 \pm 3$\,pc and is derived from parallaxes observed by the Hipparcos satellite. These results have been summarised in Table~\ref{table:summary}. It has been suggested that the two distances could be reconciled if the Pleiades cluster is metal-poor, but determinations of the atmospheric metal abundances of Pleiades F dwarfs suggest that the cluster has a solar iron abundance.

We have studied the dEB HD\,23642, a member of the Pleiades with an Hipparcos parallax determination, to calculate reliable absolute dimensions and uncertainties of the component stars. By comparing the radii of the components of HD\,23642 to theoretical models we find that the metal and helium abundances are approximately solar, which removes the possibility that the `long' and `short' distance scales could be reconciled by adopting a low metal abundance or high helium abundance for the Pleiades.

We have investigated the use of bolometric corrections (BCs) for determining the distances to eclipsing binaries, using the empirical BC calibration of Code et al. (\cite{code}) and two sources of theoretically-calculated BCs. We find that the empirical and theoretical BCs give distances to HD\,23642 in good agreement with each other. Distances determined using BCs for near-infrared filters are more precise and reliable due to a smaller dependence on interstellar reddening and metal abundance.

We have presented a new, almost entirely empirical, technique for determining the distance to dEBs composed of two components with effective temperatures between 10\,000\,K and and 3600\,K, based on interferometrically-derived calibrations between effective temperature and surface brightness (Kervella et al. \cite{KTDS}). This method does not explicitly require a light ratio for calculation of distance. Distances determined using the near-infrared $JHKL$ calibrations are more precise as uncertainties in interstellar reddening and effective temperature are less important. Using this technique and $K$-filter photometry from 2MASS, we find that HD\,23642 is at a distance of $139.1 \pm 3.5$\,pc. This distance is consistent with the `long' distance scale of the Pleiades but in disagreement with the distances to HD\,23642 and to the Pleiades derived from Hipparcos parallax observations Table~\ref{table:summary}).

Further observations of HD\,23642, to determine more accurate dimensions, would provide very precise metal abundance and effective temperature measurements for the component stars. This would reduce the uncertainty in its distance and allow further investigation of the system, which is itself an interesting object due to the metallic-lined nature of the secondary star. Further infrared observations would also allow the use of entirely empirical surface brightness relations for the calculation of an accurate distance to HD\,23642.


\begin{acknowledgements}

The authors would like to thank the referee for a very prompt and helpful response, and Dr.\ U.\ Munari for frank discussions and for making his data on HD\,23642 freely available.

The following internet-based resources were used in research for this paper: the NASA Astrophysics Data System; the SIMBAD database operated at CDS, Strasbourg, France; the VizieR service operated at CDS, Strasbourg, France; and the ar$\chi$iv scientific paper preprint service operated by Cornell University. 

JS acknowledges financial support from PPARC in the form of a postgraduate studentship. The authors acknowledge the data analysis facilities provided by the Starlink Project which is run by CCLRC on behalf of PPARC. 

This publication makes use of data products from the Two Micron All Sky Survey, which is a joint project of the University of Massachusetts and the Infrared Processing and Analysis Center/California Institute of Technology, funded by the National Aeronautics and Space Administration and the National Science Foundation. 

\end{acknowledgements}



\begin{thebibliography}{}

\bibitem[1958]{abt} Abt, H. A. 1958, ApJ, 128, 139
\bibitem[1978]{abtlevato} Abt, H. A., \& Levato, H. 1978, PASP, 90, 201
\bibitem[1991]{andersen} Andersen, J. 1991, A\&AR, 3, 91
\bibitem[1976]{BE} Barnes, T. G., \& Evans, D. S. 1976, MNRAS, 174, 489
\bibitem[1976]{BEP} Barnes, T. G., Evans, D. S., \& Parsons, S. B. 1976, MNRAS, 174, 503
\bibitem[2004]{benedict} Benedict, G. F. 2004, in IAU Coll. 196, Transit of Venus: New Views of the Solar System and Galaxy, eds. D. W. Kurtz \& G. E. Bromage, in press
\bibitem[2000]{bessell} Bessell, M. S. 2000, PASP, 112, 961
\bibitem[1998]{BCP} Bessell, M. S., Castelli, F., \& Plez, B. 1998, A\&A, 333, 231
\bibitem[1999]{beuermann} Beuermann, K., Baraffe, I., \& Hauschildt, P. 1999, A\&A, 348, 524
\bibitem[1977]{BS} Blackwell, D. E., \& Shallis, M. J. 1977, MNRAS, 180, 177
\bibitem[1990]{BF} Boesgaard, A. M., \& Friel, E. D. 1990, ApJ, 351, 467
\bibitem[2001]{carpenter} Carpenter, J. M. 2001, AJ, 121, 2851
\bibitem[2002]{CDPT} Castellani, V., Degl'Innocenti, S., Prada Moroni, P. G., \& Tordiglione, V. 2002, MNRAS, 334, 193
\bibitem[1997]{CGK} Castelli, F., Gratton, R. G., \& Kurucz, R. L. 1997, A\&A, 318, 841.
\bibitem[1995]{claret95} Claret, A. 1995, A\&AS, 109, 441
\bibitem[1997]{claret97} Claret, A. 1997, A\&AS, 125, 439
\bibitem[1998]{claret98} Claret, A. 1998, A\&AS, 131, 395
\bibitem[1995]{CG} Claret, A., \& Gim\'enez, A. 1995, A\&AS, 114, 549
\bibitem[2004]{clausen} Clausen, J. V. 2004, New Ast. Rev., 48, 679
\bibitem[1976]{code} Code, A. D., Bless, R. C., Davis, J., \& Brown, R. H. 1976, ApJ, 203, 417
\bibitem[1999]{CWH} Colavita, M. M., Wallace, J. K., Hines, B. E., et al. 1999, ApJ, 510, 505
\bibitem[1975]{crawford} Crawford, D. L. 1975, AJ, 80, 955
\bibitem[1998]{benedetto} Di Benedetto, G. P. 1998, A\&A, 339, 858
\bibitem[1997]{FG} Fouque, P., \& Gieren, W. P. 1997, A\&A, 320, 799
\bibitem[2000]{GDH} Gatewood, G., de Jonge, J. K., \& Han, I. 2000, ApJ, 533, 938
\bibitem[2002]{girardi} Girardi, L., Bertelli, G., Bressan, A., et al. 2002, A\&A, 391, 195
\bibitem[1988]{HM} Hauck, B., \& Mermilliod, M. 1998, A\&AS, 129, 431
\bibitem[1957]{johnson} Johnson, H. L. 1957, ApJ, 126, 121
\bibitem[2004]{KTDS} Kervella, P., Th\'evenin, F., Di Folco, E., \& S\'egransan D. 2004, A\&A, accepted (astro-ph/0404180) (KTDS04)
\bibitem[1988]{KTK} Kharitonov, A. V., Tereshchenko, V. M., \& Knjazeva, L. N. 1988, Alma-Ata, Nauka, p.\,484 (CDS catalogue: III/202)
\bibitem[1999]{KS} Kruszewski, A. \& Semeniuk, I. 1999, Acta Astronomica, 49, 561
\bibitem[1993]{kurucz} Kurucz, R. L. 1993, CD-ROM 13, SAO
\bibitem[1995]{KB} Kurucz, R. L., \& Bell, B. 1995, CD-ROM 23, SAO
\bibitem[1977]{lacy} Lacy, C. H. 1977, ApJ, 213, 458
\bibitem[2002]{makarov} Makarov, V. V. 2002, AJ, 124, 3299
\bibitem[1993]{MMM} Meynet, G., Mermilliod, J.-C., \& Maeder, A. 1993, A\&AS, 98, 477
\bibitem[1985]{moon} Moon T. T., 1985, Commun.\ Univ.\ London Obs.\ No.\,78
\bibitem[1985]{MD} Moon T. T., \& Dworetsky M. M, 1985, MNRAS, 217, 305.
\bibitem[2004]{M04} Munari, U., Dallaporta S., Siviero A., et al. 2004, A\&A, 418, L31 (M04)
\bibitem[1999]{NG} Narayanan, V. K., \& Gould, A. 1999, ApJ, 523, 328
\bibitem[1972]{ND} Nelson, B., \& Davis, W. D. 1972, ApJ, 174, 617
\bibitem[2004]{PSK} Pan, X., Shao, M., \& Kulkarni, S. S. 2004, Nature, 427, 326
\bibitem[1957]{pearce} Pearce, J. A. 1957, Publ.\ Dom.\ Astr.\ Obs., 10, 435
\bibitem[1997]{perryman} Perryman, M. A., Lindegren, L., Kovalevsky, J., et al. 1997, A\&A, 323, L49
\bibitem[1998]{pols} Pols, O. R., Schr\"oder, K.-P., Hurley, J. R., Tout, C. A., \& Eggleton, P. P., 1998, MNRAS, 298, 525
\bibitem[1980]{popper80} Popper, D. M. 1980, ARA\&A, 18, 115
\bibitem[1984]{popper} Popper, D. M. 1984, AJ, 89, 132
\bibitem[1981]{PE} Popper, D. M., \& Etzel, P. B. 1981, AJ, 86, 102
\bibitem[2002]{SG} Salaris, M., \& Groenewegen, M. A. T. 2002, A\&A, 381, 440
\bibitem[1994]{SS} Simon, K. P., \& Sturm, F. 1994, A\&A, 281, 286
\bibitem[1993]{smalley93} Smalley, B., 1993, MNRAS, 265, 1035
\bibitem[2004]{smalley04} Smalley, B., 2004, in IAU Symp. 224, The A-star Puzzle, eds. J. Zverko, W. W. Weiss, J. \v Zi\v z\v novsk\'y \& S. J. Adelman, in press
\bibitem[1995]{SD} Smalley, B., \& Dworetsky, M. M. 1995, A\&A, 293, 446
\bibitem[2002]{SGKB} Smalley, B., Gardiner, R. B., Kupka, F., \& Bessell, M. S. 2002, A\&A, 395, 601
\bibitem[2001]{SSD} Smalley, B., Smith, K. C., \& Dworetsky, M. M. 2001, {\sc UCLSYN} Userguide (unpublished)
\bibitem[1992]{smith} Smith, K. C. 1992, Ph.D. Thesis, University of London
\bibitem[2004a]{SMSa} Southworth, J., Maxted, P. F. L., \& Smalley, B. 2004a, MNRAS, 349, 547
\bibitem[2004b]{SMSb} Southworth, J., Maxted, P. F. L., \& Smalley, B. 2004b, MNRAS, 351, 1277
\bibitem[2004c]{SMSc} Southworth, J., Maxted, P. F. L., Smalley, B., \& Etzel P. B. 2004c, in preparation
\bibitem[1998]{SSK} Stauffer, J. R., Schultz, G., \& Kirkpatrick, J. D. 1998, ApJ, 499, L199
\bibitem[2003]{stauffer} Stauffer, J. R., Jones, B. F., Backman, D., et al. 2003, AJ, 126, 833
\bibitem[2001]{SN} Stello, D., \& Nissen, P. E. 2001, A\&A, 374, 105
\bibitem[2003]{torres} Torres, G. 2003, IBVS, 5402 
\bibitem[1999]{belle} van Belle, G. T. 1999, PASP, 111, 1515
\bibitem[1993]{hamme} van Hamme, W. 1993, AJ, 106, 2096    
\bibitem[1999]{leeuwen99} van Leeuwen, F. 1999, A\&A, 341, L71
\bibitem[2004]{leeuwen04} van Leeuwen, F. 2004, in IAU Coll. 196, Transit of Venus: New Views of the Solar System and Galaxy, eds. D. W. Kurtz \& G. E. Bromage, in press
\bibitem[1993]{wilson} Wilson, R. E., 1993, in ASP Conf.\ Ser.\ Vol.\ 38, New Frontiers in Binary Star Research, eds. Leung, K.-C., Nha, I.-S., p.\ 91
\bibitem[1971]{WD} Wilson, R. E., \& Devinney, E. J., 1971, ApJ, 166, 605
\bibitem[2004]{zwahlen} Zwahlen, N., North, P., Debernardi, Y., et al. 2004, A\&A Letter, in press (astro-ph/0408430)
\end{thebibliography}
\end{document}